# Controlling vertical magnetization shift by spin-orbit torque in ferromagnetic/antiferromagnetic/ferromagnetic heterostructure

Z. P. Zhou[1,2], X. H. Liu[1,2*], K. Y. Wang[1,2,3,4*]

[1]*State Key Laboratory for Superlattices and Microstructures, Institute of Semiconductors, Chinese Academy of Sciences, Beijing 100083, China*

[2] *Center of Materials Science and Optoelectronics Engineering, University of Chinese Academy of Sciences, Beijing 100049, China*

[3] *Beijing Academy of Quantum Information Sciences, Beijing 100193, China*

[4]*Center for Excellence in Topological Quantum Computation, University of Chinese Academy of Science, Beijing 100049, China*

\* Corresponding e-mail: xionghualiu@semi.ac.cn

kywang@semi.ac.cn

We report the control of vertical magnetization shift (VMS) and exchange bias through spin-orbit torque (SOT) in Pt/Co/Ir$_{25}$Mn$_{75}$/Co heterostructure device. The exchange bias accompanying with a large relative VMS of about 30 % is observed after applying a single pulse 40 mA in perpendicular field of 2 kOe. Furthermore, the field-free SOT-induced variations of VMS and exchange bias is also observed, which would be related to the effective built-in out-of-plane field due to unequal upward and downward interfacial spin populations. The SOT-induced switched fraction of out-of-plane interfacial spins shows a linear dependence on relative VMS, indicating the number of uncompensated pinned spins are proportional to the switched interfacial spins. Our finding offers a comprehensive understanding for electrically manipulating interfacial spins of AFM materials.

Exchange bias (EB) refers to a shift in the hysteresis loop along the magnetic field axis due to the interface exchange coupling between ferromagnetic (FM) and antiferromagnetic (AFM) materials.[1-8] This phenomenon has been widely studied because of the technological application in spintronic devices and magnetic recording.[7] Furthermore, a vertical shift of the FM magnetization is also observed in FM/AFM system, which is claimed to be due to a number of frozen uncompensated or canting AFM spins at the FM/AFM interface.[9-14] Usually, the vertical magnetization shift (VMS) in conventional FM/AFM system is quite small and cannot be easily detected by isothermal magnetization measurements, and most exchange bias models ignore the possibility of VMS.[15-17]

On the other hand, except for generating EB, AFM materials themselves have recently attracted tremendous attentions due to their abundant properties for future spintronics applications: robustness against external field, no stray fields, and ultrafast spin dynamics.[18,19] Remarkably, the discovery of electrical switching of an antiferromagnet by spin-orbit torque (SOT) motivates considerable researches in AFM spintronics.[20-24] The magnetization switching caused by SOT commonly consists of a damping-like torque $\mathbf{m} \times (\boldsymbol{\sigma} \times \mathbf{m})$ and a field-like torque $\mathbf{m} \times \boldsymbol{\sigma}$, where $\mathbf{m}$ is the spin moment and $\boldsymbol{\sigma}$ is the spin polarization of spin current.[25,26] Besides the spin Hall effect (SHE), the Rashba effect, arising from underlying crystal structure or interfaces inversion symmetry breaking, together with spin-orbit coupling also induces SOT switching.[25,26] To date, the manipulation of bulk properties of antiferromagnets has been realized by several different approaches,[18-24] the tunability of EB through SOT was recently reported by Lin et al.[27] and Liu et al.[28] Furthermore, to explore the control of VMS via SOT will give a comprehensive understanding on electrically manipulating interfacial spins of AFM materials. To realize this aim, we design the device in heavy metal (HM)/FM/AFM/FM structure, where the bottom and top FM layers are perpendicular magnetic anisotropy (PMA) and in-plane anisotropy, respectively.[29]

In this work, we report the tunability of VMS and EB by SOT in Pt(3)/Co(1)/Ir$_{25}$Mn$_{75}$(4)/Co(5) (thickness in nanometers) heterostructure. The current-induced EB with a large relative VMS of about 30 % is observed. Moreover, we demonstrated the field-free SOT-induced modifications of VMS and EB. Our work provides a route to comprehensively understand the electrical control of AFM interfacial spins at the interface of AFM and FM.

The stack structure of Ta(1)/Pt(3)/Co(1)/Ir$_{25}$Mn$_{75}$(4)/Co(5)/Ta(2) (in nanometers) was deposited on thermally oxidized Si substrate by magnetron sputtering at room temperature, as schematically illustrated in Fig. 1(a). Samples Ta(1)/Pt(3)/Co (1)/Ir$_{25}$Mn$_{75}$(4)/Ta(2) and Ta(1)/Pt(3)/Co(1.2)/Ir$_{25}$Mn$_{75}$(4)/Ru(3) without top ferromagnetic layer, and Ta(1)/Pt(3)/Co(1)/Ir$_{25}$Mn$_{75}$(4)/Co(5)/Al$_2$O$_3$(2) were studied as reference. The base pressure was less than $1 \times 10^{-8}$ Torr before deposition, and the pressure of the sputtering chamber was 0.8 mTorr during deposition. No magnetic field was applied during the sputtering. The samples were then patterned into Hall bar devices with channel width of 10 μm by photolithography and Ar-ion etching. Inset (up) of Fig. 1(b) shows the optical microscope image of the Hall bar device. All magnetic and transport properties measurements were performed at room temperature.

Figure 1(b) presents the anomalous Hall effect (AHE) resistance ($R_H$) dependence of the out-of-plane field ($H_z \perp$ film plane) for Ta(1)/Pt(3)/Co(1)/Ir$_{25}$Mn$_{75}$(4)/Co(5)/Ta(2) device before applying pulsed current, indicating the 1 nm bottom Co layer exhibits the PMA magnetization. Meanwhile, the two-step switching behavior, with a strong out-of-plane pinning due to the exchange coupling between Co(1)/IrMn(4) layers, is observed. The two-step hysteresis loops are associated with the occurrence of a bi-domain state, where the two domain populations are oppositely exchange biased because of opposite orientations of the uncompensated AFM spins at the FM/AFM interface.[30,31] The top Co layer has the magnetic easy axis in the plane [Inset (down) of Fig. 1(b) with H // film plane], the small remanence is due to a weak in-plane exchange coupling between IrMn(4)/Co(5) layers.[2,3]

The Hall bar device was then subjected to a single current pulse $I_p$ with fixed width 50 ms, in a longitudinal field $H_x$ = 1 kOe [inset (up) of Fig. 1(b)]. Through the spin Hall effect (SHE), a charge current in the ± **x** direction should produce a spin polarization along the ± **y** direction for the positive spin-Hall angle of Pt.[32-35] The resulting spin current can switch the magnetization of bottom PMA Co between the ± **z** directions, provided that both the current density and $H_x$ are large enough. Moreover, the absorption of transverse spin currents is reported to vary with the FM thickness with a characteristic saturation length of 1.2 nm.[36] Thus in our device, not only the 1 nm thick Co layer but also the AFM interfacial spins can be directly affected by SOT, leading to the variation of EB.[28] $R_H$ vs. $H_z$ loops, obtained after applying a single current pulse $I_p$ = 40 mA and -40 mA in $H_x$ = 1 kOe, are shown in Figs. 1(c) and 1(d), respectively. In this process, firstly, we set $H_x$ = 1 kOe, and then applied a single pulse $I_p$ = 40 mA; after that we set $H_x$ = 0 and $I_p$ = 0, and measured the $R_H$ vs. $H_z$ loop. The $R_H$ vs. $H_z$ loop for $I_p$ = -40 mA under $H_x$ = 1 kOe is obtained by using the same process. The main part of the loop shows positive EB for 40 mA [Fig. 1(c)] and negative EB for -40 mA [Fig. 1(d)], demonstrating the interfacial spins of Co(1)/IrMn(4) can be switched by SOT.[28] The $R_H$ vs. $H_z$ loops, obtained after applying different single current pulses in $H_x$ = 1 kOe are displayed in Fig. S1 of supplementary material.

Furthermore, utilizing a single pulse 40 mA under $H_z$ = 2 kOe generates a single-step $R_H$ vs. $H_z$ loop with positive EB [Fig. 1(e)], suggesting a complete alignment of the interfacial spins of Co(1)/IrMn(4) in the direction of $H_z$. Strikingly, the $R_H$ vs. $H_z$ loop indeed displays an obviously vertical magnetization shift (VMS) as we expect, which is not observed for the device after applying $I_p$ in $H_x$ [Figs. 1(c)-(d)]. The exchange bias field $H_E$ and VMS $R_{HE}$ are marked in Fig. 1(e), and the relative VMS, defined as $|R_{HE}/R_{HS+}|$, reaches about 30 % [The $R_{HS+}$ and $R_{HS-}$ are saturated $R_H$ at positive (2200 Oe) and negative (-2200 Oe) field, respectively], which is much larger than that reported in conventional field-cooling AFM/FM system.[9-12] As the positive and negative $I_p$ have nearly the same effect on switching out-of-plane interfacial spins under $H_z$,[28] we measured the device using a single pulse 40 mA under $H_z$ = -2 kOe [Fig. 1(f)], which exhibits a negative EB but the same VMS as that in Fig. 1(e).

As a comparison, the VMS is not observed in Ta(1)/Pt(3)/Co(1)/IrMn(4)/Ta(2) device (Supplementary Fig. S2). It has been reported that the VMS observed in FM/AFM system is mainly due to a number of frozen uncompensated or canting AFM spins at the FM/AFM interface.[9-14] For Ta(1)/Pt(3)/Co(1)/IrMn(4)/Ta(2) device, the interfacial exchange coupling takes place at Co(1)/IrMn(4), it is very difficult to froze uncompensated or cant AFM spins in one direction at room temperature. However, when covering a top Co layer (for 4 nm IrMn, the interlayer exchange coupling between two FM layers can be ignored.[37]), the interfacial exchange couplings in Co(1)/IrMn(4) (out-of-plane) and in IrMn(4)/Co(5) (in-plane) modify the micromagnetic structure of IrMn.[38] After using $I_p$ under $H_z$, partial uncompensated AFM spins are pinned in **z** direction, resulting in the VMS. The same VMS for the device after $I_p$ in positive and negative $H_z$ would be related to its special AFM microstructure. Whereas, for the device applying $I_p$ under $H_x$, there is no preferred pinned AFM spins in **z** direction, we thus cannot observe the VMS [Figs. 1(c)-(d) and Fig. S1].

In addition, we found modifications of both EB and VMS induced by SOT in zero field [Figs. 1(h)-(j)]. The initial state was set by applying a single pulse 40 mA under $H_z$ = 2 kOe [Fig. 1(g)], and subsequent hysteresis loops were obtained after utilizing each single pulse $I_p$ under zero field. For $I_p$ = 26 mA, the single-step $R_H$ vs. $H_z$ loop moves up [Fig. 1(h)] as compared to the case in Fig. 1(g) (see the dashed line). With increasing $I_p$ to 30 mA, the out-of-plane interfacial spins of Co(1)/IrMn(4) are also varied by SOT, corresponding to a clear step (as the arrow indicates) in $R_H$ vs. $H_z$

loop [Fig. 1(i)]. The switched fraction of out-of-plane interfacial spins, defined as the ratio of the switched $R_H$ step height to the whole loop's height ($R_{HS+}$ - $R_{HS-}$), is about 30 %. When $I_p$ increases to 40 mA, the shape of two-step $R_H$ vs. $H_z$ curve [Fig. 1(j)] becomes similar to the as-deposited condition [Fig. 1(b)].

The relative VMS ($|R_{HE}/R_{HS+}|$) and the switched fraction of out-of-plane interfacial spins dependence of $I_p$ at zero field, starting from an initial state set by applying a single pulse $I_p$ = 40 mA under $H_z$ = 2 kOe, are summarized in Figs. 2(a) and 2(b), respectively. With increasing $I_p$ from 28 to 32 mA, the $|R_{HE}/R_{HS+}|$ gradually decreases from 12 % to about 0 [Fig. 2(a)] while the switched fraction increases from 3 % to the saturated value of 43 % [Fig. 2(b)]. Importantly, the $|R_{HE}/R_{HS+}|$ linearly varies with switched fraction [Fig. 2(c)].

To understand the reason for the field-free SOT-induced variations of EB and VMS, it necessary to consider the effect of antiferromagnetic domain structure of the IrMn. The formation of inequivalent upward and downward domain populations after applying a single pulse $I_p$ = 40 mA under $H_z$ = 2 kOe, generates an effective out-of-plane field ($H_{z\text{-eff}}$). The built-in $H_{z\text{-eff}}$ helps the SOT to alter the out-of-plane interfacial spins from a metastable domain state to an equilibrium state.[28] With the decrease of $H_{z\text{-eff}}$, the frozen AFM spins (corresponding VMS) gradually reduces and the switched faction of interfacial spins gradually increases [Fig. 2(c)]. As $H_{z\text{-eff}}$ disappears for equilibrium state, the out-of-plane interfacial spins couldn't be modified by SOT without external field.[32-35] As a result, the VMS disappears and the switched faction of interfacial spins becomes constant, as shown in Figs. 2(a) and (b).

Importantly, we in fact found an efficient way to initialize the interfacial spins via SOT at zero field. Moreover, we excluded the role of Joule heating in the observed EB and VMS (Supplementary Figs. S3 and S4),[39] and the spin current contribution is mainly from Pt (Supplementary Fig. S5).

From the Meiklejohn and Bean model,[1-3] the macroscopic $H_E$ of fully uncompensated and pinned FM/AFM interface can be written as a function of the unidirectional magnetic interface energy σ and the Heisenberg-like interface exchange energy $J$, $H_E = \dfrac{\sigma}{M_{FM} t_{FM}} = J \dfrac{S_{AFM} S_{FM}}{a_{AFM}^2 M_{FM} t_{FM}}$ (1) where $M_{FM}$ and $t_{FM}$ are the magnetization and the thickness of the FM, $a_{AFM}$ is the size of the unit cell of the AFM, and $S_{AFM}$ and $S_{FM}$ are the spins of the interfacial AFM and FM atoms. Since $J$, $M_{FM}$, $t_{FM}$ and $a_{AFM}$ are constant, the variation of $H_E$ associating with the magnetization reversal of the bottom Co by SOT, corresponding to the switched fraction, mainly reveals the change of $S_{AFM}$. The linear relationship between |$R_{HE}/R_{HS+}$| and switched fraction suggests that the number of uncompensated pinned AFM spins is inversely proportional to the switched $S_{AFM}$.[11,12] The maximum (minimum) |$R_{HE}/R_{HS+}$| corresponds to the minimum (maximum) switched fraction via SOT at zero field [Fig. 2(c)].

To obtain more insight into the control of VMS via SOT in HM/FM/AFM/FM device, we measured the $R_H$ vs. $H_z$ loops using a constant single pulse $I_p$ = 29 mA under different $H_z$, starting from an initial state set by applying a single pulse 40 mA under $H_z$ = 2 kOe [Fig. 3]. Clearly, the changes of both EB and VMS are observed. The |$R_{HE}/R_{HS+}$| and switched fraction of out-of-plane interfacial spins versus $H_z$ at $I_p$ =

29 mA are plotted in Figs. 4(a) and 4(b), respectively. Two distinct behaviors are observed: $|R_{HE}/R_{HS+}|$ slightly decreases with enhancing $H_z$ from -10 Oe and reaches a minimum of 5 % for -200 Oe, then this value gradually increases to about 30 % for -2000 Oe. Correspondingly, the switched fraction sharply increases from 26 % for -10 Oe to 80 % for -200 Oe, and then slightly increases to 100 % with further increasing $H_z$.

It is noteworthy that the curves in Figs. 4(a)-(b) can be divided into two parts, as the dashed line indicates at $H_z$ = -200 Oe. The linear variation of $|R_{HE}/R_{HS+}|$ with switched fraction exhibit opposite sign of slope for these two parts [Figs. 4(c)-4(d)]. Unlike the zero field case, the VMS always exists for all the $H_z$ range [Fig. 4(a)]. In this process, both the external field $H_z$ and built-in $H_{z\text{-eff}}$ take effect in variation of $|R_{HE}/R_{HS+}|$ and switched fraction but the $H_z$ plays dominant role. For $|H_z|$ < 200 Oe, the trend of $|R_{HE}/R_{HS+}|$ with switched fraction is similar to the zero field case. Here the $H_z$ takes two effects: (i) assists the out-of-plane interfacial spins rapid switch and (ii) prevents the full relaxation of uncompensated pinned spins to zero, through SOT. When $|H_z|$ > 200 Oe, a slight enhancement of the switched fraction leads to a great increase in $|R_{HE}/R_{HS+}|$, accordingly the larger $H_z$ induces more uncompensated out-of-plane pinned spins via SOT.

The formation of the two regions separated at $H_z$ = -200 Oe in Fig. 4(a) would be related to the AFM domain structure of the IrMn. For our Pt/Co/IrMn/Co device, partial AFM domains would not toward **z**, which leads to the pinning of a part of the interfacial spins in directions deviated from **z** direction. The interfacial spins with effective spins moments along in-plane cannot be changed to **z** direction via SOT under $H_x$ [see a clear step in $R_H$ vs. $H_z$ curves in Figs. 1(c) and (d)]. Whereas, the SOT under large $H_z$ can flip all the spins to **z** direction with the help of strong Zeeman interaction. Hence the sharp increase of $|R_{HE}/R_{HS+}|$ in $|H_z| > 200$ Oe might be due to the rotation of in-plane pinning by IrMn domains to **z** direction. Therefore, we implemented the control of VMS and EB via SOT in Pt/Co/Ir$_{25}$Mn$_{75}$/Co heterostructure.

In conclusion, we studied the manipulation of VMS and EB through SOT in Pt(3)/Co(1)/Ir$_{25}$Mn$_{75}$(4)/Co(5) heterostructure device. The current-induced EB with a large relative VMS of about 30 % is found. Furthermore, the field-free SOT-induced variations of VMS and EB is also observed, which would be associated with the effective built-in $H_{z\text{-eff}}$ due to unequal upward and downward interfacial spin populations. We demonstrated an effective way to initialize the interfacial spins through SOT at zero field. The linear relationship between switched fraction of out-of-plane interfacial spins with $|R_{HE}/R_{HS+}|$ reveals that the number of uncompensated pinned spins are proportional to the switched interfacial spins by SOT. Our work would be very important for understanding and utilizing the AFM interfacial spins in related spintronic applications.

See supplementary material for EB change by different single pulses in $H_x = 1$ kOe for device Ta(1)/Pt(3)/Co(1)/IrMn(4)/Co(5)/Ta(2), EB change by a single pulse $I_p = 26$ mA in $H_z = \pm 2$ kOe for sample Ta(1)/Pt(3)/Co(1)/IrMn(4)/Ta(2), determinations of the temperature increase due to Joule heating during the single current pulse and the blocking temperature of the IrMn layer for sample Ta(1)/Pt(3)/Co(1)/IrMn(4)/Ta(2), and the variation of EB by SOT in Ta(1)/Pt(3)/Co(1)/IrMn(4)/Co(5)/Al$_2$O$_3$(2) and Ta(1)/Pt(3)/Co(1.2)/IrMn(4)/Ru(3) devices.

This work was supported by the National Key R&D Program of China No. 2017YFB0405700. This work was also supported by the NSFC Grant No.s 11474272 and 61774144 and Beijing Natural Science Foundation Key Program Grant No. Z190007. The Project was sponsored by the Chinese Academy of Sciences, Grant No.s QYZDY-SSW-JSC020, XDB28000000, and XDPB12 as well.

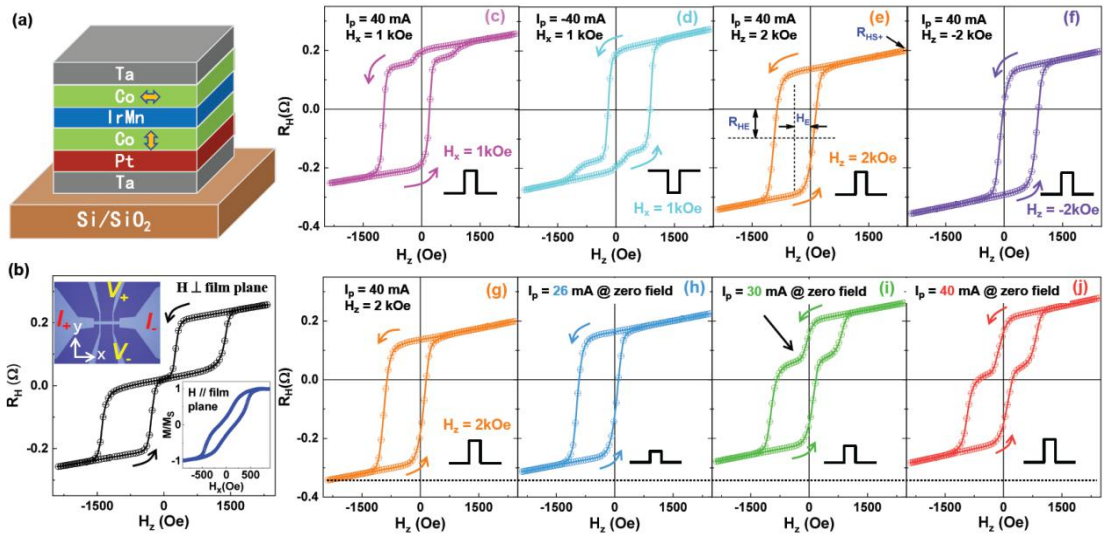

**Fig. 1**. (a) Schematic of the stack, orange double headed arrows indicate the easy axis of the two ferromagnetic layers. (b) $R_H$ vs. $H_z$ curve for sample Ta(1)/Pt(3)/Co(1)/Ir$_{25}$Mn$_{75}$(4)/Co(5)/Ta(2) before applying current pulses. Inset: (up) Optical micrograph of the fabricated Hall device and measurement scheme with the definition of x-y-z coordinates. (down) Magnetization hysteresis loop (H // film plane) by a MOKE magnetometer. $R_H$ vs. $H_z$ curves after applying a single pulse 40 mA (c) or -40 mA (d) in $H_x$ = 1 kOe. $R_H$ vs. $H_z$ curves after utilizing a single pulse 40 mA in $H_z$ = 2 kOe (e) or -2 kOe (f). For the initial state set by applying $I_p$ = 40 mA under $H_z$ = 2 kOe (g), $R_H$ vs. $H_z$ curves were obtained after using each single pulse $I_p$ = 26 mA (h), 30 mA (i) and 40 mA (j) in zero field, respectively.

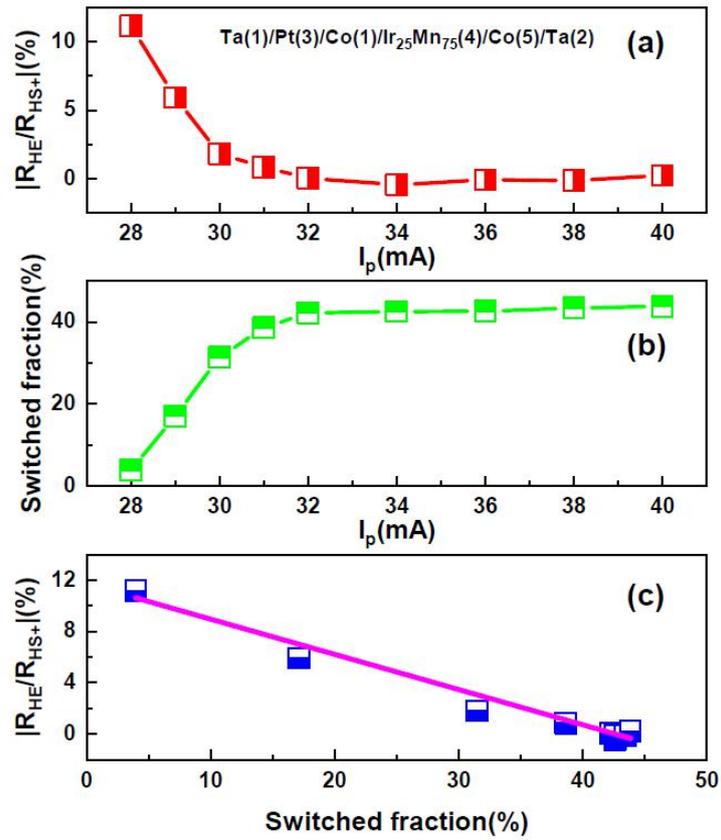

**Fig. 2**. (a) The relative VMS |$R_{HE}/R_{HS+}$| and (b) the switched fraction of out-of-plane interfacial spins versus pulsed current. (c) The |$R_{HE}/R_{HS+}$| linearly varies with switched faction.

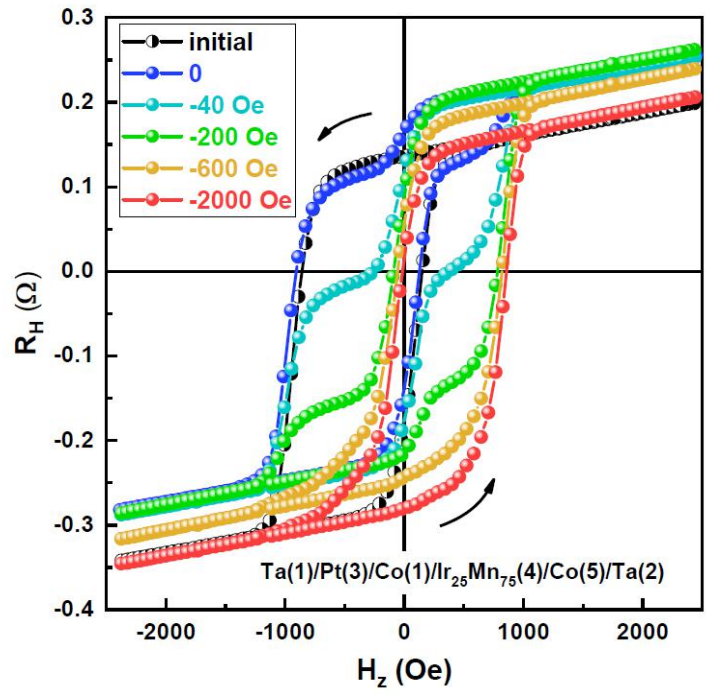

**Fig. 3**. $R_H$ vs. $H_z$ curves after applying a single current pulse $I_p$ = 29 mA under different $H_z$ (0, -40, -200, -600 and -2000 Oe).

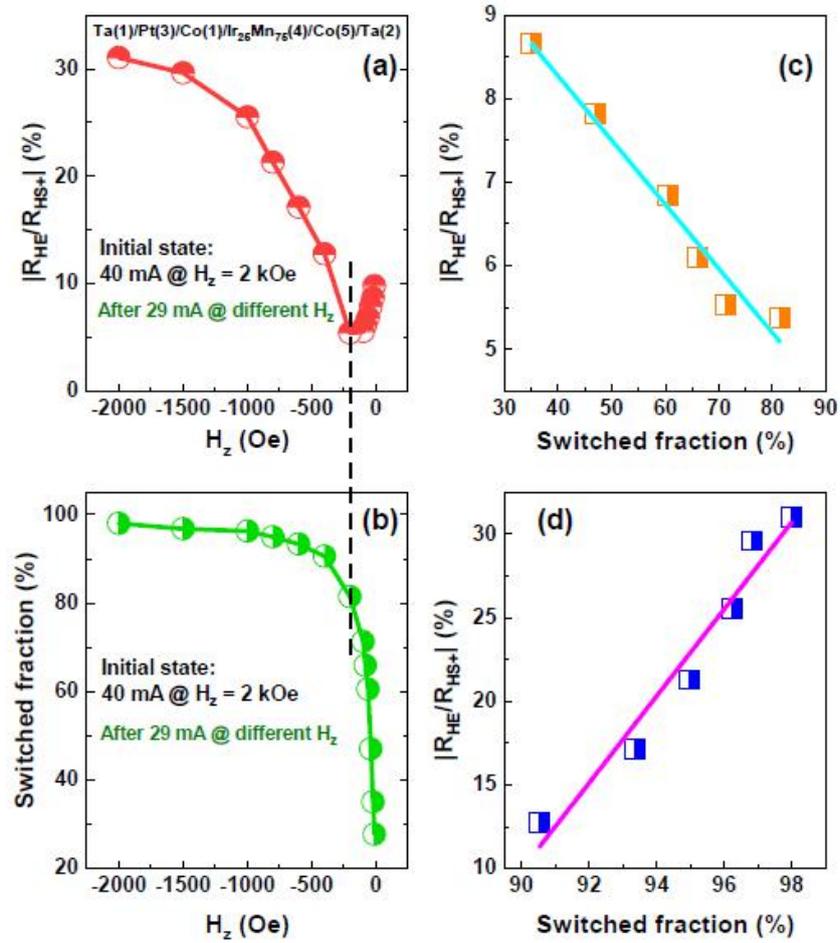

**Fig. 4.**(a) The relative VMS $|R_{HE}/R_{HS+}|$ and (b) the switched fraction of out-of-plane interfacial spins versus $H_z$. (c), (d) The linear relationship between $|R_{HE}/R_{HS+}|$ and switched fraction of out-of-plane interfacial spins for the two $H_z$ ranges separated by the dashed line in (a) and (b).

**Supplementary Information**

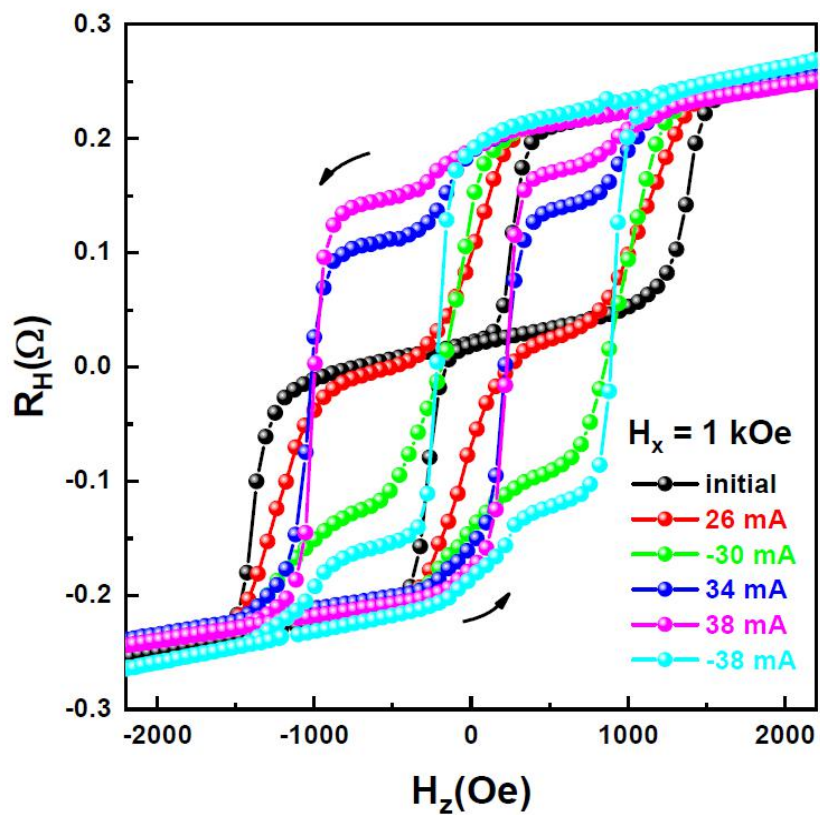

**Fig. S1**. $R_H$ vs. $H_z$ curves after applying different single pulses $I_p$ = 26 mA, -30 mA, 34 mA, 38 mA and -38 mA in $H_x$ = 1 kOe for sample Ta (1)/Pt (3)/Co (1)/IrMn (4)/Co (5)/Ta (2).

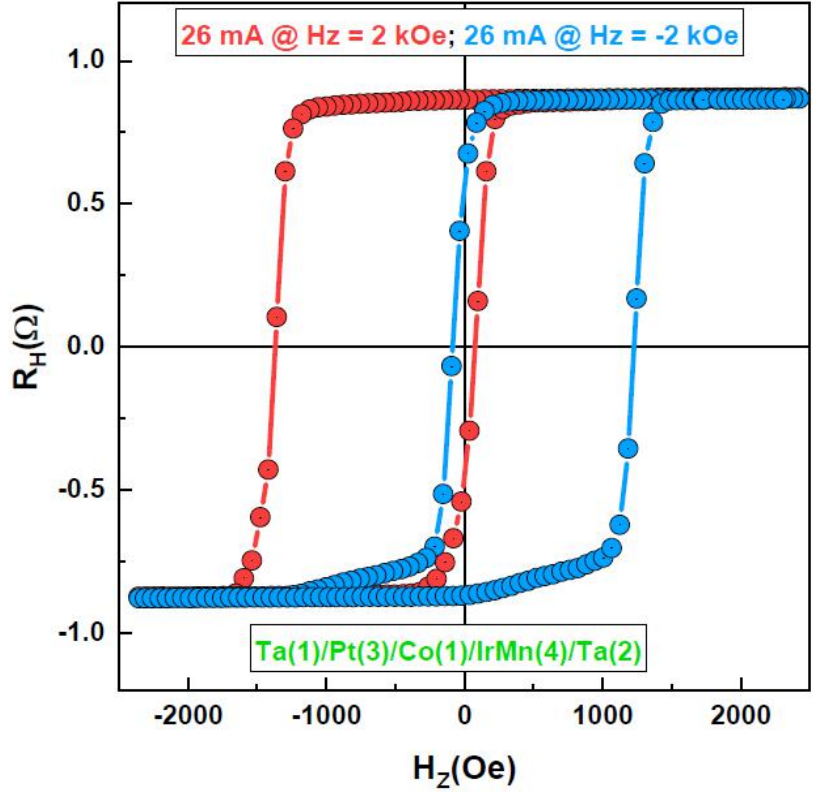

**Fig. S2**. $R_H$ vs. $H_z$ curves after applying a single pulse $I_p$ = 26 mA in $H_z$ = 2 kOe and -2 kOe for sample Ta (1)/Pt (3)/Co (1)/IrMn (4)/Ta (2). No vertical magnetization shift is observed.

To check whether the observed variation of VMS and EB is caused by current-induced Joule heating over the blocking temperature of IrMn, we designed the experiment to estimate the temperature rise. The resistance of the sample was measured during the current pulse for the sample Ta(1)/Pt(3)/Co(1)/Ir$_{25}$Mn$_{75}$(4)/Ta(2) [The EB switching can be observed after using a single pulse $I_p$ = 26 mA under $H_z$ = ± 2 kOe in Supplementary Fig. S2]. By comparing this to the measured temperature-dependence of resistance, a temperature rise of around 34 K was estimated for a single pulse $I_p$ = 30 mA with 50 ms (see Fig. S3). On the other hand, the blocking temperature for the Ta(1)/Pt(3)/Co(1)/Ir$_{25}$Mn$_{75}$(4)/Ta(2) sample, defined as the temperature where the EB disappears, is around 460 K (see Fig. S4). Therefore, we exclude a significant role of Joule heating in the observed EB and VMS variations.

Furthermore, IrMn alloys were reported to have a spin Hall angle with the same sign as that of Pt but with smaller value.[1,2] From the SOT switching data, the dominant contribution is from the bottom Pt layer, moreover, the resistivity of IrMn is about one order bigger than that of Pt, so that the current density in the Pt layer is about ten times larger. Thus the spin current contribution from IrMn can be ignored in this system, similar to the work in Refs. [3,4]. To exclude the influence of Ta, we made new devices Ta (1)/Pt (3)/Co (1)/IrMn (4)/Co (5)/Al$_2$O$_3$ (2) and Ta (1)/Pt (3)/Co (1.2)/IrMn (4)/Ru (3). Similarly, obvious VMS can be observed in the former device but no VMS can be seen in the latter one (see Fig. S5).

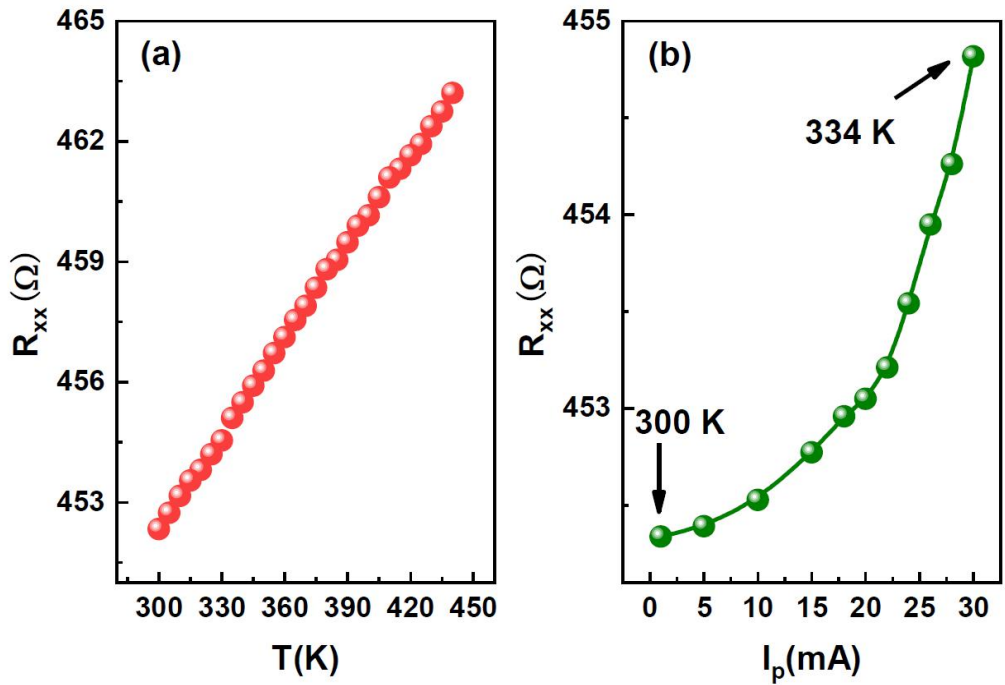

**Fig. S3.** Determination of the temperature increase due to Joule heating during the single current pulse for sample Ta (1)/Pt (3)/Co (1)/IrMn (4)/Ta (2). (a) Resistance versus temperature. (b) Resistance measured during the single current pulse, versus the current pulse magnitude $I_p$ for fixed pulse width of 50 ms.

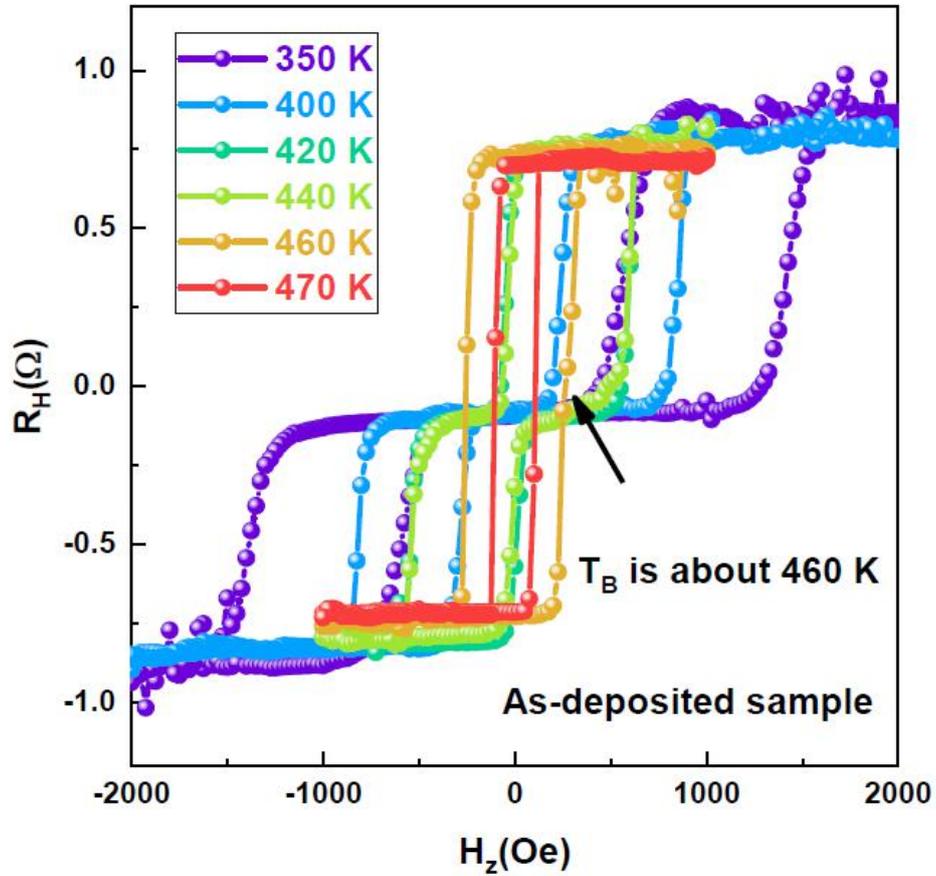

**Fig. S4**. Determination of the blocking temperature of the IrMn layer for sample Ta (1)/Pt (3)/Co (1)/IrMn (4)/Ta (2). $R_H$ vs. $H_z$ loops at different temperatures, with increasing temperature from 300 K. The two-step switching of $R_H$ vs. $H_z$ loop nearly disappears at around 460 K (as the arrow indicates), corresponding to the blocking temperature of 4 nm IrMn.

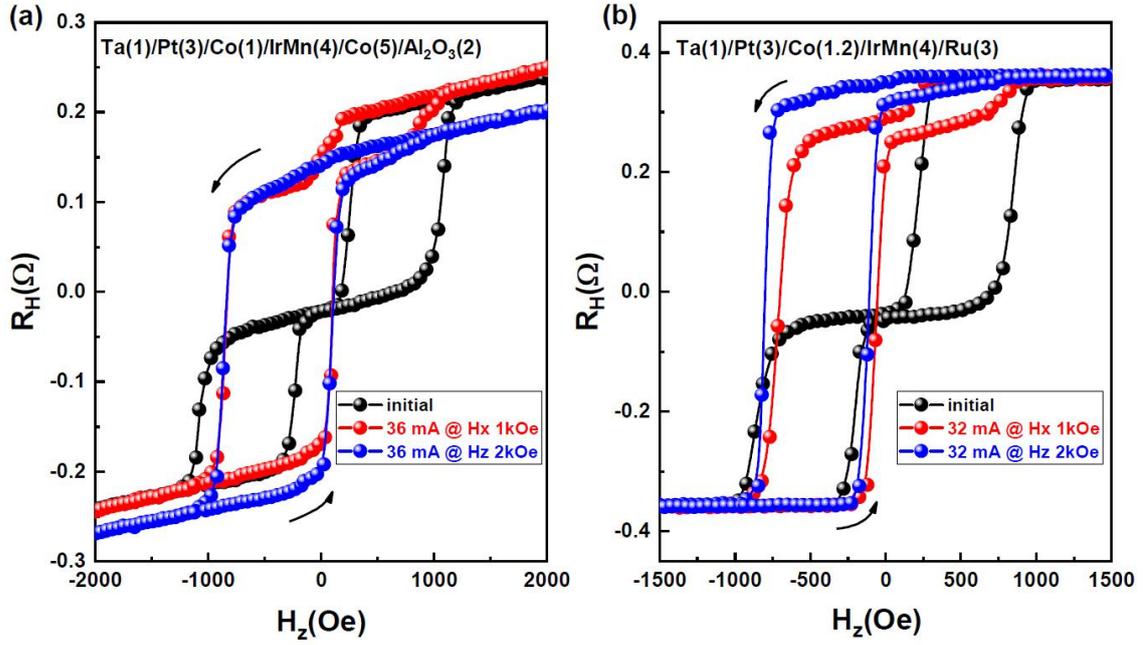

**Fig. S5**. (a) $R_H$ vs. $H_z$ curves for initial (as-grown), and after applying a single pulse $I_p$ = 36 mA in $H_x$ = 1 kOe or $H_z$ = 2 kOe for sample Ta (1)/Pt (3)/Co (1)/IrMn (4)/Co (5)/Al$_2$O$_3$ (2). (b) $R_H$ vs. $H_z$ curves for initial (as-grown), and after applying a single pulse $I_p$ = 32 mA in $H_x$ = 1 kOe or $H_z$ = 2 kOe for sample Ta (1)/Pt (3)/Co (1.2)/IrMn (4)/Ru (3).